\documentclass[aps,prl,twocolumn,amssymb,showpacs]{revtex4}  
\usepackage[dvips]{graphicx} 
\usepackage{bm}    
\usepackage{amssymb}
\newcommand{\cv}{{\bm c}}    
\newcommand{\xv}{{\bm x}}    
\newcommand{\uv}{{\bm u}}    
\newcommand{\vv}{{\bm v}}    
    
\begin{document}    
\title{Lattice Boltzmann method at finite-Knudsen numbers}
\author{Federico Toschi and Sauro Succi} 
\affiliation{Istituto Applicazioni Calcolo, CNR, V.le del Policlinico
  137, 00161, Roma, Italy }
\date{\today}
%\pacs{47.11.+j}{Computational methods in fluid dynamics}
%\pacs{47.15.-x}{Laminar flows} \pacs{67.40.Hf}{Hydrodynamics in
%  specific geometries, flow in narrow channels}
\begin{abstract}    
  A modified lattice Boltzmann model with a stochastic relaxation
  mechanism mimicking ``virtual'' collisions between free-streaming
  particles and solid walls is introduced.  This modified scheme
  permits to compute plane channel flows in satisfactory 
  agreement with analytical results over a broad spectrum of Knudsen
  numbers, ranging from the hydrodynamic regime, all the way to
  quasi-free flow regimes up to $Kn \sim 30$.
%\baselineskip 8ex    
%SUSUB\baselineskip 8ex   
\end{abstract}    
\maketitle
      
The dynamic behaviour of flows far from hydrodynamic equilibrium is an
important subject of non-equilibrium thermodynamics, with many
applications in science and engineering.  The non-hydrodynamic regime
is characterized by strong departures from local equilibrium which are
hardly handled on analytical means. Consequently, much work is being
devoted to the development of computational techniques capable of
dealing with the aforementioned non-perturbative and non-local
effects. Recently, the lattice Boltzmann (LB) method has attracted
considerable interest as an alternative to the discretization of the
Navier-Stokes equations for the numerical simulation of a variety of
complex flows \cite{LBE}. Extending LB methods to non-hydrodynamic
regimes is a conceptual challenge on its own, with a variety of
microfluidic applications, such as flows in micro and nano
electro-mechanical devices (NEMS, MEMS) \cite{CHI}.  Departures from
local equilibrium are measured by the Knudsen number, namely the ratio
of molecular mean free path, $l_m$, to the shortest hydrodynamic
scale, $l_h$: $Kn=l_m/l_h$.  Ordinary fluids feature $Kn<0.01$, while
high Knudsen numbers are typically associated to rarefied gas
dynamics, where $l_m$ is large because density is small, typical case
being aero-astronautics applications.  More recently, however, the
finite-Kn regime is becoming more and more relevant for a variety of
microfluidics applications in which the Knudsen number is large
because of the increasingly smaller size of the devices.  It is also
worth emphasizing that the finite-Knudsen regime may also bear
relevance to the problem of modeling fluid turbulence \cite{EU,SCI}.
Recent work indicates that LB may offer quantitatively correct
information also in the finite-Kn regime \cite{NIE}. This hints at the
possibility that LB may complement or even replace expensive
microscopic simulation techniques such as kinetic Monte Carlo and/or
molecular dynamics.  On the other hand, this is also puzzling because
finite-Knudsen flows are in principle expected to receive substantial
contributions from high-order kinetic moments, whose dynamics is
arguably not quantitatively correct because of lack of symmetry of the
discrete lattice \cite{LL,DELLAR}.  In this work we point out that,
irrespectively of symmetry considerations, any LB extension aiming at
describing non-hydrodynamic flows in the finite-Knudsen regime must
first provide an adequate description of the dynamics of
free-streamers, namely ballistic particles whose motion, in the limit
of infinite Knudsen number, escapes any thermalization by either bulk
or wall collisions.
 
This problem is connected to a pathology of the LB model, and in fact,
of {\it any} discrete velocity model representing a whole set of
molecular speeds within a finite solid angle, by a single discrete
speed.  The practical consequence of this drastic reduction of degrees
of freedom in momentum space is a very efficient numerical scheme, but
also the pathology connected to the fact that particles moving
parallel to the wall (for example in a channel flow) never impact on the
boundaries, so that under the effect of pressure gradients or body
forces within the fluid flow, they may enter an unrealistic
free-acceleration (runaway) regime.
%SS
More generally, runaway must be expected whenever the particle mean
free path exceeds the longest free-flight distance allowed by the
geometrical set up.
%SS
    
\begin{figure} 
\begin{center}   
  \includegraphics[width=.66\hsize]{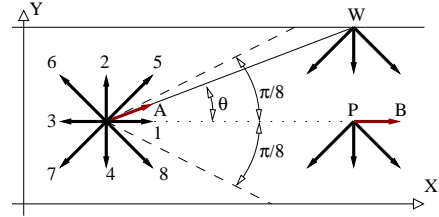}  
\end{center}  
\caption{A simplified sketch showing the velocity structure for the D2Q9    
  model in a plane channel flow.  In LBE models an hypothetic particle
  forming an angle $\theta$ with respect to the $\hat x$ axis
  (population ``A'') is counted amongst the particles in population
  ``1''. This particle would then, in the absence of collisions (i.e.
  for high-Knudsen numbers), be forced to stream parallel to the $\hat
  x$ axis without ever impacting on the walls.  In real flows instead,
  its natural fate would be to move along a straight line until a
  collision with the wall would occurr at position ``W''.}
\label{fig:pato}    
\end{figure}    
    
In this work we propose a mechanism to obviate this runaway pathology.
Our method is demonstrated through the numerical simulations of a
force-driven plane Poiseuille flow, as compared with existing
analytical theories \cite{CERCI}.
    
%\section{The Lattice Boltzmann method}    
    
The lattice Boltzmann method has been described in detail in many
publications, and here we shall only remind a few basic ideas behind
this technique.  The simplest lattice Boltzmann equation looks as
follows \cite{LBGK}:
\begin{eqnarray}    
f_i(\xv +\Delta t \cv_i,t+\Delta t)-f_i(\xv,t)=\\    
\nonumber -\omega \Delta t [f_i(\xv,t) -f_i^e(\xv,t)] + F_i \Delta t    
\label{LBGK}    
\end{eqnarray}    
where $f_i(\xv,t) \equiv f(\xv,\vv=\cv_i,t)$, $i=1, \dots, n$, is the
probability of finding a particle at lattice site $\xv$ at time $t$,
moving along the lattice direction defined by the discrete speed
$\cv_i$, and $\Delta t$ is the time step.  The left-hand side of this
equation represents free-streaming, while the right-hand side
describes collisions via a simple relaxation towards local equilibrium
$f_i^e$ (a local Maxwellian expanded to second order in the fluid
speed) in a time lapse $\tau = \omega^{-1}$.  This relaxation time
fixes the fluid kinematic viscosity as $\nu=c_s^2(\tau-1/2)$, where
$c_s$ is the sound-speed of the lattice fluid, ($c_s=1/\sqrt{3}$ in
the present work).  Finally, $F_i$ represents the effects of an
external force.  The set of discrete speeds must be chosen such that
mass, momentum and energy conservation are fulfilled, together with
rotational symmetry.  Once these symmetries are secured, the fluid
density $\rho=\sum_i f_i$, and speed $\uv = \sum_i f_i \cv_i/\rho$ can
be shown to evolve according to the (quasi-incompressible)
Navier-Stokes equations of fluid-dynamics.  In this paper, we shall
refer to the nine-speed D2Q9 model shown in Figure \ref{fig:pato}.
    
%\section{Boundary conditions}    
 
While at low Knudsen local equilibrium is established via bulk
collisions, in the regime $Kn \ge 1$, the leading equilibration
mechanism is provided by the interaction with the boundaries.
Various types of boundary conditions have been used for LB simulations
of finite-Knudsen flows \cite{BOON,SLIP,KWOK,LIM}.  In this work, we
confine our attention to bounce back (BB) \cite{BB} and kinetic, as
recently introduced by Ansumali and Karlin (AK) \cite{AK}.  The former
implement no-slip flow speed at the wall via particle reflections,
while the latter are based on the idea of reinjecting particles from
the wall according to a local equilibrium distribution with wall speed
and temperature.  While AK were explicitly designed with
finite-Knudsen effects in mind, this is clearly not the case for BB.
At high-Knudsen, neither BB nor AK are expected to provide realistic
results, since neither of them 
has been designed to do so.

%has means to tame the runaway problem
%previously discussed.  

It is therefore of interest to compare these
two quite distinct types of boundary conditions 
throughout the full range of Knudsen numbers.
   
%\subsection{Lattice Boltzmann with virtual wall collisions}    
    
In a intermediate/high-Knudsen flow the main mechanism for
thermalisation is provided by collisions with solid walls.  This
mechanism is non-local, meaning by this that, in order to be
thermalized, a particle residing at a given lattice site (see
population ``A'' in Fig.  \ref{fig:pato}) needs to reach the wall
(position ``W'' in Fig.  \ref{fig:pato}), where it is scattered along
some different direction, thereby transferring some momentum to the
wall.  From Figure \ref{fig:pato} it is clear that in the LB
representation, all molecules with momentum in the solid angle
$[-\pi/8:\pi/8]$ around population ``1'', are collapsed into
population ``1''.  In the limit of infinite Knudsen numbers,
population ``1'', being parallel to the walls, behaves like a
ballistic beam, whose motion escapes both bulk collisions and wall
scattering. Due to the presence of an external drive, these beam
particles enter a free-acceleration regime, which ultimately leads to
runaway behaviour in time.  Therefore, in the way to a finite-Knudsen
LB scheme, no matter how accurate in terms of tensorial symmetry, a
mechanism to thermalize these ballistic streamers needs to be
introduced.
    
Here we propose a mechanism to mimick non-local wall collisions
through {\it virtual wall collisions} (VWC for short). At every time
step each free-streamer is assigned a {\it virtual speed} (called
``A'' in Fig. \ref{fig:pato}) $(v_x,v_y)=c\cdot (\cos \theta,
\sin\theta)$, with $\theta$ uniformly distributed in $[-\pi/8,
\pi/8]$. Assuming collisions are distributed according to a
Poissonian, we define $p$ as the probability to ``hit'' (i.e. to
collide at least once) the wall in a single time step, multiplied by
the probability that no molecular collisions occurred during that time
step. In practice, this means that a wall-collision event is performed
at each time step and each lattice site with probability:
\begin{equation}    
p(x,y;t) = e^{-1/Kn} \cdot \left(1 - e^{-{{c\, dt\, \sin(\theta(x,y))} \over H}} \right)    
\end{equation}    
where the first factor is the probability of undergoing no bulk
collision during the flight of (average) length $H$ to the wall, when
the average mean-free-path for bulk collisions is $\lambda$ (i.e.
$-1/Kn = - \lambda/H$).  The second term describes the fact that after
a flight of length $L$ the particle hits the wall with probability
one. This is modeled at each time step in the following way.  In a
time step $dt$ the particle moves a distance $dl=c\,dt$; since on
average there is one collision every $L$ lattice spacings, the
probability of a wall collision in a time step $dt$ is given by
$\exp{(-dl/L)} = \exp{(-c\,dt\,\sin(\theta)/H)}$ (in our units $c\, dt
= 1$).
    
Note that $p$ goes to zero in the limit of $Kn \rightarrow 0$, so that  
virtual wall collisions fade away in the hydrodynamic regime, as they  
should. Due to the non-analytic dependence in $Kn$ the hydrodynamic  
limit is recovered faster than any power of $Kn$.  
   
The probability $p$ goes to zero also in the limit $\theta \rightarrow  
0$, corresponding to the case of the virtual speed falling back into a  
free-streamer (see Figure 1). However, by design, the occurrence of  
such a limit has now a vanishingly small probability, much like in a  
continuum flow.  
   
Upon hitting the wall, the free-streamer feeds the span-wise moving  
populations according to the rules adopted for true wall-collisions  
(see Figure 1).  For instance, for a virtual collision of a  
right-moving particle with the top wall, we have:  
\[    
f'_1 = f_1 (1-p),\;\;f'_{7,8}=f_{7,8}+pf_1/6,\;\;    
f'_4 = f_{4}+2pf_1/3\;    
\]    
where the coefficients $\{1/6,4/6,1/6\}$ correspond to the weights of the  
nine-speed local equilibrium.  
    
Thanks to this mechanism, momentum is removed from populations aligned
with the wall boundaries and re-distributed along the direction
orthogonal to the boundaries. This allows the flow to relax towards a
(non-local) equilibrium also in the limit of high $Kn$ numbers.
    
%\section{Simulation conditions}    
    
We simulate a two-dimensional flow in a rectangular duct of size $L$
along the streamwise direction ($x$) and width $H<L$ along $y$.  The
flow is driven by a volumetric force 
(force per unit volume) $F_x=\rho c \frac{8 \nu U_0}{H^2}$
along the streamwise direction.  In the hydrodynamic limit, such force
leads to a parabolic velocity profile of amplitude $U_0$.  Both BB and
AK boundary conditions are applied on-site.  The Knudsen number is
tuned by changing the value of the viscosity, according to the
expression:
\[    
Kn = l_m/H  = \nu /(c_s H)    
\]    
Knudsen numbers in the range $10^{-3}<Kn<30$ have been simulated, at a
fixed Mach number $Ma=u/c_s=0.03$.  The grid resolution used was $101
\times 21$, although a few simulations at higher resolution of $101
\times 41$ have also been performed to check the grid-independence of
our results. After discarding the initial transient time, velocity
profiles were averaged in time 
%SS
in order to minimize statistical
fluctuations associated with the stochastic nature of the
VWC mechanism..  
%SS
The simulation was run until the
relative change of the profiles fell below a given threshold
(comparable with machine precision).
   
\begin{figure}    
\begin{center}
  \includegraphics[width=.66\hsize]{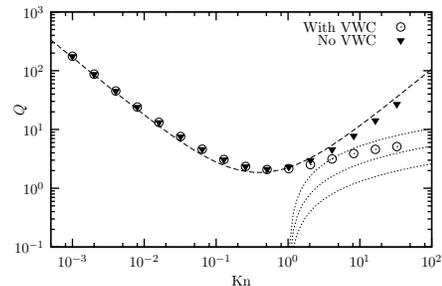} 
\end{center}   
\caption{AK boundary contions: normalized mass flux as a function     
  of $Kn$ number, with ($\circ$) and without ($\blacktriangledown$)
  virtual wall collisions.  The dashed line is Cercignani's prediction
  (\ref{CERCI1}) for low-$Kn$. Dotted lines are prediction
  (\ref{CERCI2}) for high-$Kn$ multiplied by factors $1, 2$ and $4$,
  from bottom to top.  Following Cercignani, the mass flow is
  normalized with $6 \rho \nu$.}
\label{FIG2}    
\end{figure}

\begin{figure}    
\begin{center}
\includegraphics[width=.66\hsize]{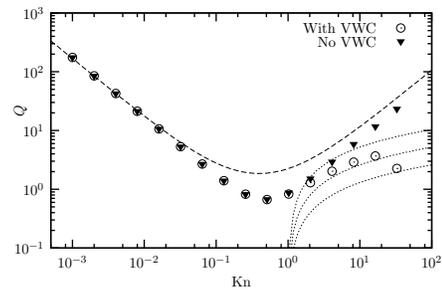} 
\end{center}     
\caption{BB boundary contions: normalized mass flux as a function     
  of $Kn$ number, with ($\circ$) and without ($\blacktriangledown$)
  virtual wall collisions.  Notations are the same as in the Figure
  \ref{FIG2}.}.
\label{FIG2b}    
\end{figure}

%\subsection{Mass flow rate}    
One of the major successes in the early days of kinetic theory was the
prediction of a minimum of the mass flow rate as a function of the
Knudsen number around $Kn \sim 1$.  Interestingly, both BB and AK
boundary conditions exhibit a minimum mass flow rate around $Kn \sim
1$ (see Figures \ref{FIG2} and \ref{FIG2b}).  This indicates that,
contrary to the common tenet, LB can capture the so-called ``Knudsen
paradox''.  However, they both severely overestimate the mass flow in
the high-Kn region, $Kn>1$.  This is due to the presence of a finite
fraction of ballistic streamers, as discussed earlier on.  Figures
\ref{FIG2} and \ref{FIG2b} show the behaviour of the normalized mass
flow rate $Q=\sum_y u_x(y)/(6 \nu)$, for the case with and without
virtual wall collisions.
   
The numerical results are compared with the analytical expression
obtained by Cercignani in the zero and infinite Knudsen limits
respectively (both fluxes are normalized in units of $6 \rho \nu$):
\begin{eqnarray}    
\label{CERCI1}    
Q_0&=&1/(6 Kn)+s+(2s^2-1)\;Kn\\  
\label{CERCI2}  Q_{\infty} &\sim& \pi^{-1/2} \; \log{(Kn)}    
\end{eqnarray}    
where $s=1.015$ (see \cite{CERCI}).    
    
By increasing the $Kn$ number the flow slips more and more, until the
centerline speed becomes comparable with the sound speed, so that the
LBE simulation breaks down.
  
As shown in Figure \ref{FIG2}, by introducing virtual wall collisions,
a nearly quantitative agreement with analytical results is observed.
It is worth stressing that no free parameter has been used in the
simulations.  The highest achievable $Kn$ number may not be high
enough to test the asymptotic prediction. 
Therefore, in order to appreciate the
quality of our results, we have plotted Cercignani's prediction,
$Q_{\infty}$, in Figures \ref{FIG2} and \ref{FIG2b}, together with its
magnifications by factors $2$ and $4$.

Back in 1867, Maxwell predicted that the slip length of a rarefied gas
flowing on a semi-infinite slab should be proportional to the mean
free path, $l_s=1.146 Kn$, \cite{MAX}.  Nearly four decades ago,
Cercignani computed the next order corrections in $Kn$, leading to the
following quadratic expression for the slip speed:
\begin{equation}    
\label{CERCI}    
V_s = a Kn + b Kn^2    
\end{equation}
where $a=1.146\cdot \left({{\partial u} \over {\partial
      y}}\right)_{y=-H/2}$ and $b=-0.9075\cdot \left({{\partial^2 u}
    \over {\partial y}^2}\right)_{y=-H/2}$.
    
\begin{figure}
\begin{center}
  \includegraphics[width=.66\hsize]{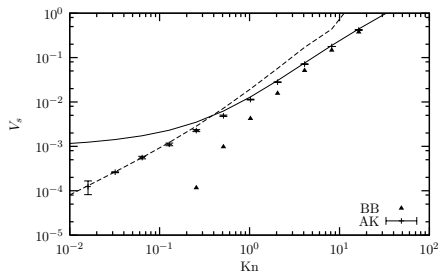}    
\end{center}
\caption{Wall slip velocity as a function of the Knudsen number for     
  AK (plusses) and BB (triangles) boundary conditions.  The solid line
  is the wall slip velocity extrapolated from the values of the flux
  computed assuming a flat velocity distribution across the channel.
  The dashed line is Cercignani's low-Knudsen analytical prediction
  (\ref{CERCI}).  Error bars as described in the text.}
\label{FIG3}    
\end{figure}

The dependence of the slip speed on the Knudsen number is presented in
Figure \ref{FIG3}.  The slip velocity is defined as the value (at the
wall position) of the parabolic fit of the averaged velocity profiles.
Boundary data were excluded from the fit.  The error bars on the slip
velocity were estimated as the discrepancy between the results of the
$21\times 101$ and $41\times 101$ simulations.
  
From Figure \ref{FIG3} it is clearly seen that BB boundary conditions
fail to reproduce the slip velocity.  Thus, our results support
previous criticism on the use of BB boundary conditions at
finite-Knudsen \cite{LUO,AK}.  However, kinetic boundary conditions
provide quantitative agreement with analytical results up to $Kn \sim
1$, even beyond the theoretical expectations.  At intermediate and
high Knudsen both BB and AK provide similarly inaccurate results. This
is because the physics becomes more and more insensitive to the BB or
AK boundary conditions and dependent on the VWC mechanism instead.
Indeed, as virtual collisions are introduced, both BB and AK provide
satisfactory agremeent with analytical data.  The solid line in Figure
\ref{FIG3} corresponds to the ratio $Q/(\rho H)$ and the fact that
this quantity matches almost exactly the measured slip speed indicates
that the flow profile is basically flat, as predicted by the
asymptotic theory.  Finally, our results do not extend below $Kn=0.01$
simply because an accurate evaluation of the very small values of
$V_s$ in this regime requires a much higher transverse resolution than
adopted here.
   
%%%%%%%%%%%%%%%%    
%%%%%%%%%%%%%%%%    
    
%Some concluding remarks are in order.  It is sometimes argued that the
%use of a 9-speed LB for non hydrodynamic regimes is doomed because
%this model does not have enough symmetry to accomodate a
%quantitatively correct evolution of the higher order fields, including
%the kinetic tensor.  This work shows that, prior to any symmetry
%considerations, thermalization of free-streamers has to be addressed
%first, because even higher order lattices would still suffer from the
%runaway problem.  In this paper we have presented a simple, but
%effective, procedure to handle this problem.  
Our results indicate that a standard nine-speed LB scheme equipped
with Ansumali-Karlin boundary conditions {\it and} a virtual wall
collision mechanism, can capture salient features of channel flow in
both hydrodynamic and strongly non-hydrodynamic regimes.  Of course,
this is only the first step towards a systematic inclusion of
high-Knudsen effects in the lattice kinetic framework, and much
further work is needed to address more general situations, such as
non ideal geometries, high shear rates \cite{TRO} and thermal effects
\cite{GARCIA}.  Another interesting point to be explored for the
future is the potential benefit of using multi-relaxation
\cite{HSB,MTR} and entropic LB \cite{ELB} schemes for a better
description of fluid-wall interactions.
   
%\subsection{Acknowledgments}     
The authors thank S. Ansumali and I. Karlin for discussions and for
making available their implementation of AK boundary conditions in a
early stage of this project.  SS acknowledge discussions with A.
Garcia.  Financial support from the NATO Collaborative Link Grant
PST.CLG.976357 and CNR Grant (CNRC00BCBF-001) is acknowledged.

\end{document}